\documentclass[prd,twocolumn,showpacs,amsfonts,amssymb,amsmath, nofootinbib]{revtex4-1}

\usepackage{graphicx}
\usepackage{simplewick}
\usepackage{enumitem}
\usepackage{epstopdf}
\usepackage{multirow}
\usepackage{dcolumn}
\usepackage{hyperref}
\usepackage{cases}
\usepackage{bm}
\usepackage{epsfig}
\usepackage{color}
\usepackage[normalem]{ulem}

\def\be{\begin{equation}}
\def\ee{\end{equation}}
\def\bea{\begin{eqnarray}}
\def\eea{\end{eqnarray}}

\newcommand{\bear}{\begin{eqnarray}}

\newcommand{\eear}{\end{eqnarray}}

\newlength{\tskip}
\newbox\pippobox

\def\be{\begin{equation}}
\def\ee{\end{equation}}
\def\bea{\begin{eqnarray}}
\def\eea{\end{eqnarray}}
 
\newcommand{\nver}{\hat{\mathbf{n}}}
\newcommand{\kver}{\hat{\mathbf{k}}}

\begin{document}

\title{Kinetic Sunyaev-Zel'dovich effect in modified gravity}

\author{Federico Bianchini$^{1,2}$}
\email{corresponding author: fbianchini@sissa.it}
\author{Alessandra Silvestri$^{3}$}
\affiliation{
\smallskip
$^{1}$ SISSA, International School for Advanced Studies, Via Bonomea 265, 34136, Trieste, Italy \\
\smallskip
$^{2}$ INFN, Sezione di Trieste, Via Valerio 2, I-34127 Trieste, Italy\\
$^{3}$ Institute Lorentz, Leiden University, PO Box 9506, Leiden 2300 RA, The Netherlands \\
\smallskip }

\begin{abstract}
We investigate the impact of modified theories of gravity on the kinetic Sunyaev-Zel'dovich (kSZ) effect of the cosmic microwave background. We focus on a  specific class of $f(R)$ models of gravity and compare their predictions for the kSZ power spectrum to that of the $\Lambda$CDM model. We use a publicly available modified version of Halofit to properly include the nonlinear matter power spectrum of $f(R)$  in the modeling of the kSZ signal. We find that the well-known modifications of the growth rate of structure in $f(R)$ can indeed induce sizable  changes in the kSZ signal, which are more significant than the changes induced by modifications of the expansion history. We discuss prospects of using the kSZ signal as a complementary probe of modified gravity, giving an overview of assumptions and possible caveats in the modeling. 
\end{abstract}


\maketitle

%
\section{Introduction}\label{Sec:Intro}
The journey of the photons from the last scattering surface to us is not a smooth one. On their way to the observer, they undergo several physical processes which induce additional anisotropies on top of the primordial ones, imprinted at the time of recombination.  These can be noticed as  secondary effects on the Cosmic Microwave Background (CMB), of which  the integrated Sachs-Wolfe effect and CMB lensing are common examples.  Since they happen at different intermediate redshifts, they carry valuable information on the Universe between the last scattering and today. In this paper we focus on the kinetic Sunyaev-Zel'dovich (kSZ) effect, which is sourced by the inverse Compton scattering of CMB photons off clouds of moving free electrons~\cite{1980MNRAS.190..413S,1987ApJ...322..597V, Jaffe:1997ye,Shaw:2011sy}. The kSZ signal is sensitive to both the expansion history and the dynamics of cosmological perturbations. Its constraining power as a geometrical probe was investigated in~\cite{Ma:2013taq}. Here we focus on the growth of structure and  investigate the prospect of using the kSZ effect to constrain models of modified gravity that address the phenomenon of cosmic acceleration. Indeed we show that the kSZ phenomenon is particularly sensitive to the growth of perturbations at redshifts $z\lesssim 2$ and, as such, we expect it to be a good probe of cosmological models that introduce late times modifications of the growth rate of structure, such as clustering dark energy and modified gravity. In particular,  we consider the class of $f(R)$ models introduced by Hu and Sawicki in~\cite{Hu:2007nk} and we find that the corresponding modifications to the growth rate of structure can indeed induce sizable changes in the kSZ effect. We  discuss also the caveats and assumptions associated to the modeling and measurements of the kSZ signal, and conclude giving an outlook of the observational prospective. \\
A wide set of observational probes has been proposed and exploited to test theories of modified gravity from astrophysical to cosmological scales. In particular, viable $f(R)$ models have been constrained with secondary CMB anisotropies, such as CMB lensing, the integrated Sachs-Wolfe effect \citep{Raveri:2014cka,Ade:2015rim} and its cross-correlation with galaxy density \citep{Song:2007da,giannantonio10}, galaxy clusters abundances \citep{Jain:2007yk, Ferraro:2010gh, Cataneo:2014kaa} and profiles \citep{lombriser12,Wilcox:2015kna}, galaxy power spectrum \citep{Oyaizu:2008tb,Dossett:2014oia}, redshift-space distortions from spectroscopic surveys \citep{Guzzo:2008ac,yamamoto10,Alam:2015rsa}, weak gravitational lensing \citep{Simpson:2012ra, Harnois-Deraps:2015ula}, 21-cm intensity mapping \citep{hall13}, and dwarf galaxies \citep{Jain:2012tn,Vikram:2013uba}.\\
 The paper is organized as follows. In Sec.~\ref{Sec:kSZ} we provide an overview of the kSZ effect and describe our modeling of the kSZ angular power spectrum. The different gravity models considered in this analysis are reviewed in Sec.~\ref{Sec:Models}.  We investigate the effect of the modified cosmic growth history on the kSZ observable and discuss possible caveats in Sec.~\ref{Sec:Results}. In Sec.~\ref{Sec:kSZ_obs} we discuss the detectability of such an effect with the available and upcoming observations, while the conclusions are presented in Sec.~\ref{Sec:concl}. Throughout the paper we assume a fiducial background expansion that closely mimics one of a spatially flat $\Lambda$CDM cosmological model consistent with the best-fit cosmological parameters derived using the joint data sets of 2015 Planck TT + low P \citep{Ade:2015xua}, namely $\Omega_b=0.049$, $\Omega_c=0.264$, $\Omega_{\Lambda}=0.687$, $n_s=0.966$, $H_0=67.31$ km s$^{-1}$ Mpc$^{-1}$, and $z_{\rm re}=9.9$.

\section{The Kinetic Sunyaev-Zel'dovich effect}\label{Sec:kSZ}
In this section we review the physics and the modeling of the kSZ signal.  During their journey from the last scattering surface to the observer, CMB photons experience further Compton scattering off clouds of moving free electrons: the line-of-sight (LOS) component of the electron momentum induces temperature fluctuations in the CMB sky through Doppler effect. This phenomenon is known as kinetic Sunyaev-Zel'dovich effect and is given by~\citep{1980MNRAS.190..413S}
\be
\label{delta_T_v}
\frac{\Delta T}{T_0}(\nver)  = \sigma_T \int \frac{dz }{(1+z)H(z)}n_e(z) e^{-\tau(z)} \mathbf{v} \cdot \nver,
\ee
where $\sigma_T$ is the Thomson scattering cross section, $T_0\simeq 2.725$ K is the average CMB temperature, $H(z)$, $\tau(z)$, $n_e(z)$, and $\mathbf{v} \cdot \nver$ are the Hubble parameter, optical depth, free electron number density, and peculiar velocity component along the LOS. \\
The optical depth out to redshift $z$ is 
\be
\tau(z) = \sigma_T c \int_0^{z} dz \frac{\bar{n}_e(z)}{(1+z)H(z)},
\ee
 where $\bar{n}_e$ is the mean free-electron density,
\be
\bar{n}_e = \frac{\chi\rho_g(z)}{\mu_e m_p},
\ee
with $\rho_g(z) = \rho_{g0}(1+z)^3$ being the mean gas density at redshift $z$ and $\mu_em_p$ the mean mass per electron. The electron ionization fraction is defined as~\citep{Shaw:2011sy,Ma:2013taq}
\be
\label{chi}
\chi = \frac{1-Y_{\rm p}(1-N_{\rm He}/4)}{1-Y_{\rm p}/2}.
\ee
Here, $Y_{\rm p}$ is the primordial helium abundance, fixed to $Y_{\rm p} = 0.24$, and $N_{\rm He}$ is the number of helium electrons ionized.

The kSZ power is expected to be sourced by a contribution from the epoch of reionization, dubbed the patchy kSZ,  and one from the postreionization epoch, namely the homogeneous kSZ. The former is originated by inhomogeneities in the electron density and the ionization fraction at redshift $z \gtrsim z_{\rm re}$ \footnote{We define $z_{\rm re}$ as the redshift at which hydrogen reionization ends.}; its power spectrum amplitude and shape depend, at first order, on the time and duration of reionization~\citep{2005ApJ...630..643M,2005ApJ...630..657Z}. The latter is sourced by free-electron density perturbations and peculiar velocities at $z \lesssim z_{\rm re}$. Since we focus on the  kSZ signal after reionization, we set the upper integration limit in Eq.~\ref{delta_T_v} to $z_{\rm re} = 9.9$ while we assume $\chi=0.86$, i.e. neutral helium at all redshifts \footnote{This is an assumption: in a more realistic case, helium atoms remain singly ionized until $z\sim 3$ ($\chi=0.93$) and, below that, they are thought to be doubly ionized ($\chi = 1$). \cite{Shaw:2011sy} calculates for their $\Lambda$CDM case that this would increase the kSZ power by a factor of 1.22 at $\ell = 3000$ relative to their baseline model. Hence, the level of uncertainty on the kSZ power spectrum due to helium reionization is equivalent to that due to the uncertainty on $\sigma_8$.}. We discuss more in detail about the possible choices for the reionization redshift and the dependence of kSZ on its value in Sec.~\ref{Sec:caveats}.

Writing $n_e = \bar{n}_e(1+\delta)$, Eq.~\ref{delta_T_v} can be recast as 
\be
\label{delta_T_q}
\frac{\Delta T}{T_0}(\nver)  = \Bigl( \frac{\sigma_T\rho_{g0}}{\mu_e m_p} \Bigr) \int_0^{z_{\rm re}} dz \frac{(1+z)^2}{H(z)}\chi e^{-\tau(z)} \mathbf{q} \cdot \nver,
\ee
where $\mathbf{q} = \mathbf{v}(1+\delta)$ is the density weighted peculiar velocity (ionized electron peculiar momentum).\\

Since the longitudinal Fourier modes of $\mathbf{q}$, i.e those  with $\mathbf{k}$ parallel to $\nver$, experience several cancellations in the LOS integral of Eq.~\ref{delta_T_q}, only transverse momentum modes contribute to the effect~\citep{1987ApJ...322..597V,Jaffe:1997ye}. Moreover, in the linear regime the velocity field is purely longitudinal, so that only the cross term $\mathbf{v}\delta$ can source the kSZ power spectrum. In the small angle limit, the kSZ angular power spectrum can be written under the Limber approximation~\citep{1953ApJ...117..134L} as
\be
\label{cl_ksz}
\begin{split}
C_{\ell} &= \frac{8\pi^2}{(2\ell+1)^3} \Bigl( \frac{\sigma_T\rho_{g0}}{\mu_e m_p} \Bigr)^2  \\ &\times\int_0^{z_{re}} \frac{dz}{c} (1+z)^4 \chi^2 \Delta^2_B(\ell/x,z) e^{-2\tau(z)}\frac{x(z)}{H(z)},
\end{split} 
\ee
where $x(z)=\int_0^z(cdz'/H(z'))$ is the comoving distance at redshift z, $k=\ell/x$ and $\Delta_B(k,z)$  is the power spectrum of the transverse (curl) component of the momentum field. Earlier analytical studies calculated the expression for $\Delta^2_B$~\citep{1987ApJ...322..597V,Jaffe:1997ye,1995ApJ...439..503D,Ma:2001xr} to be
\be
\label{delta_B_full}
\begin{split}
\Delta^2_B(k,z)&= \frac{k^3}{2\pi^2} \int \frac{d^3\mathbf{k}'}{(2\pi)^3}(1-\mu^2)\Bigl[P_{\delta\delta}(|\mathbf{k}-\mathbf{k'}|, z) P_{vv}(k',z)  \\
& - \frac{k'}{|\mathbf{k}-\mathbf{k'}|} P_{\delta v}(|\mathbf{k}-\mathbf{k'}|,z) P_{\delta v}(k',z) \Bigr],
\end{split} 
\ee
where $P_{\delta\delta}$ and $P_{vv}$ are the linear matter density and velocity power spectra, while $P_{\delta v}$ is the density-velocity cross spectrum and $\mu=\kver \cdot \kver'$. We can further simplify Eq.~\ref{delta_B_full} by relating the peculiar velocity field to the density perturbations: in the linear regime and on subhorizon scales, under the assumption that $\dot{\Phi}\simeq 0$, we can use the continuity equation for matter to write $\tilde{\mathbf{v}} = i f \dot{a} \tilde{\delta} \mathbf{k}/k^2$, where $f(a,k)=d\log\delta(a,k)/d\log a$ is the linear growth rate. This gives us the following equations
\be
\label{}
P_{vv}(k) = \Bigl(\frac{f\dot{a}}{k} \Bigr)^2 P_{\delta\delta}(k); \qquad P_{\delta v}(k) = \Bigl(\frac{f\dot{a}}{k} \Bigr) P_{\delta\delta}(k).
\ee
Inserting them into Eq.~\ref{delta_B_full}, we obtain\footnote{Redshift dependence is suppressed for clarity.}
\be
\label{delta_B_lin}
\Delta^2_B(k)= \frac{k^3}{2\pi^2}\dot{a}^2 \int \frac{d^3\mathbf{k}'}{(2\pi)^3}f^2(k') P_{\delta\delta}(|\mathbf{k}-\mathbf{k'}|) P_{\delta\delta}(k')I(k,k'), 
\ee
where 
\be
\label{I_kk}
I(k,k') = \frac{k(k-2k'\mu)(1-\mu^2)}{k'^2(k^2+k'^2-2kk'\mu)}
\ee
is the kernel that describes how linear density and velocity fields couple to each other. Since we are interested in studying the impact of scale-dependent growth of structure induced by modified gravity scenarios on the kSZ observable, $f$ is kept inside the previous integral.
Combining Eq.~\ref{cl_ksz} with Eq.~\ref{delta_B_lin}, one obtains the so-called Ostriker-Vishniac (OV) effect~\citep{1986ApJ...306L..51O}.

When nonlinear structure formation arises, linear perturbation theory does not hold and Eq.~\ref{delta_B_lin} breaks down: several previous works have investigated the impact of nonlinearities on the kSZ power spectrum~\citep{Hu:1999vq,Ma:2001xr,Zhang:2003nr,Shaw:2011sy}. In particular,~\cite{Hu:1999vq,Ma:2001xr} argue that nonlinearities in the velocity field are suppressed by a factor of $1/k^2$ with respect to those in the density field. Hence it is sufficient to  replace the linear $P_{\delta\delta}(k,z)$ in Eq.~(\ref{delta_B_lin}) with its nonlinear counterpart $P^{\rm NL}_{\delta\delta}(k,z)$ to capture the effect of nonlinearities, so that
\begin{equation}
\label{delta_B_nl}
\resizebox{\columnwidth}{!} 
{
    $ \Delta^2_B(k)= \frac{k^3}{2\pi^2}\dot{a}^2 \int \frac{d^3\mathbf{k}'}{(2\pi)^3}f^2(k') P^{\rm NL}_{\delta\delta}(|\mathbf{k}-\mathbf{k'}|) P_{\delta\delta}(k')I(k,k'). $
}
\end{equation}

Throughout this work, we use Eq.~\ref{delta_B_nl} and Eq.~\ref{cl_ksz} to compute theoretical predictions for the full kSZ angular power spectra. Let us notice that the continuity equation that we have used to go from (\ref{delta_B_full}) to (\ref{delta_B_lin}) remains valid in our Jordan frame treatment of $f(R)$ gravity. N-body simulations have shown that $f(R)$ models affect significantly the velocity spectrum, with resulting changes in $P_{vv}$ with respect to the $\Lambda$CDM being more sizable than the corresponding changes in $P_{\delta\delta}$~\cite{Li:2012by}. Nevertheless, the relative size of these modifications should not hinder the above assumption on negligibility of nonlinearities of the velocity power spectrum in $\Delta_B^2$.  

As previously done~\citep{Jaffe:1997ye,1995ApJ...439..503D,Ma:2001xr,Zhang:2003nr}, we assume that gas fluctuations trace dark matter ones at all scales, i.e. there is no (velocity) bias between them. Note that this approximation breaks on small scales, where baryon thermal pressure tends to make gas distribution less clustered than the dark matter one. This suppression effect due to baryon physics can be incorporated into a window function $W(k,z)$, such that~\cite{Shaw:2011sy}
\be
\label{gas_window}
P^{\rm NL}_{\rm gas}(k,z) =  W^2(k,z)P^{\rm NL}_{\rm DM}(k,z).
\ee

An investigation of the astrophysical processes such as radiative cooling, star formation, and supernova feedback on the kSZ power spectrum achieved by measuring the window function $W^2(k,z)$ in hydrodynamic simulations is provided in~\cite{Shaw:2011sy} (CSF model). There, the authors provide an improved fitting formula for $W^2(k,z)$ with respect to the one presented in~\cite{1998MNRAS.296...44G} and argue that the CSF model is a robust lower bound on the homogeneous kSZ amplitude.

All the previous studies on which we rely to model the kSZ signal, neglect the contribution of the connected term in the transverse component of the momentum power spectrum\footnote{As we have seen, $\mathbf{q}\approx \mathbf{v}\delta$, hence $\langle qq \rangle = \langle vv \rangle \langle \delta\delta \rangle + 2\langle v\delta \rangle^2 + \langle v\delta v\delta \rangle_c$} that arises in the nonlinear regime: a recent investigation of this term is reported in~\cite{2015arXiv150605177P}.

Since we aim at studying the modified gravity effects on the kSZ power spectrum, in the following analysis we  adopt the full nonlinear kSZ modeling without including the thermal gas pressure. However, when comparing with observations, one should take that into account and exploit numerical simulations to calibrate. As discussed in~\cite{Shaw:2011sy}, the realistic expected homogeneous kSZ power spectrum should lie within the region delimitated by the OV signal (lower bound) and the full kSZ (upper bound).

\section{Models}\label{Sec:Models}
In our analysis we compare the kSZ effect in the standard cosmological model, $\Lambda$CDM to that  in the $f(R)$ class of modified gravity. As we discuss, the latter models have an expansion history which is very close to the $\Lambda$CDM one, while they can predict a different cosmic growth history. In our analysis we therefore focus on the effects of the modified growth.  

For our standard gravity scenario, we consider a flat $\Lambda$CDM composed of baryonic matter (with critical density $\Omega_b$), pressureless cold dark matter ($\Omega_{c}$), and a cosmological constant $\Lambda$ exerting a negative pressure and dominating the current energy budget of the Universe ($\Omega_{\Lambda}$). 
The power spectrum of the transverse component of the momentum field  $\Delta^2_B(k,z)$~[Eq.~\ref{delta_B_nl}], which sources the kSZ signal, is sensitive to the growth history  through the linear growth rate and to the matter power spectrum, both the linear and nonlinear one. 
In our analysis, we use $f(z)=\Omega_m(z)^\gamma$~\cite{Linder:2005in} , where  $\gamma \simeq 0.55$ for general relativity, to evaluate the growth rate, while we make use of CAMB\footnote{\url{http://cosmologist.info/camb/}}~\citep{camb}  to compute the matter power spectrum. In this way, nonlinearities in $P^{\rm NL}_{\delta\delta}(k)$ are taken into account via the Halofit prescription of~\cite{halofit}.

For the case of $f(R)$ gravity~\cite{Song:2006ej,Bean:2006up,Pogosian:2007sw,DeFelice:2010aj}, we  consider a specific family of  models introduced by Hu and Sawicki in~\cite{Hu:2007nk}. This represents one of the viable $f(R)$ families, capable of sourcing the late-time  accelerated expansion, closely mimicking the $\Lambda$CDM background cosmology on large scales and evading solar system tests because of the built-in chameleon mechanism~\citep{Hu:2007nk,Khoury:2003rn}. The action of the model in the Jordan frame reads
\be
\label{fR_action}
S=\frac{1}{16\pi G}\int d^4x \sqrt{-g}[R+f(R)]+S_m,
\ee
where $f(R)$ is a generic function of the Ricci scalar
\be
\label{fR_HS}
f(R) = - m^2 \frac{c_1(-R/m^2)^n}{c_2(-R/m^2)^n+1},
\ee
and the matter sector is minimally coupled to gravity. Here, $c_1$, $c_2$, and $n$ are free dimensionless parameters of the theory and $m^2$ is a mass scale. Because of the higher order derivative nature of the theory, there is a dynamical scalar degree of freedom $f_R \equiv df/dR$, dubbed the scalaron, that mediates a long-range fifth-force and is responsible for the modification of structure formation. 

Following~\cite{Hu:2007nk}, we fix the mass scale to $m^2=8\pi G\bar{\rho}_0/3$, where $\bar{\rho}_0$ is the average density of matter today. This effectively corresponds to having $m^2/R\ll 1$ for the entire expansion history, allowing an expansion of~(\ref{fR_HS}) in $m^2/R$, with the scalaron sitting always close to the minimum of the potential, and the model resembling, at linear order, the $\Lambda$CDM one.
If we additionally fix $c_1/c_2\sim6\Omega_\Lambda/\Omega_m$, then in first approximation  the expansion history mimics closely that of  a $\Lambda$CDM universe with a cosmological constant $\Omega_\Lambda$ and matter density $\Omega_m$. This leaves us with two free parameters, $c_1/c_2^2$ and $n$; models with larger $n$ mimic $\Lambda$CDM until later in cosmic time, while models with smaller $c_1/c_2^2$ mimic it more closely.

In the following we work in terms of $n$ and the present value of the scalaron $f^0_R$, which can be related to $c_1/c_2^2$ as~\citep{Hu:2007nk}
\be
\frac{c_1}{c_2^2} = -\frac{1}{n}\Biggl[ 3 \Bigl( 1 + 4\frac{\Omega_{\Lambda}}{\Omega_m} \Bigr) \Biggr]^{n+1} f^0_R.
\ee
In order to calculate the kSZ power spectrum, we  use the publicly available MGCAMB~\cite{Zhao:2008bn} and MGHalofit\cite{mghalofit}   to compute the growth rate and the linear and nonlinear matter power spectrum. Following the discussion above, we approximate the expansion history to the $\Lambda$CDM one, and focus on differences due to a different growth rate of structure.  MGCAMB and MGHalofit use the quasistatic approximation for the dynamics of scalar perturbations, which is sufficiently good for the choice of parameters that we have made above and for the range of $f^0_R$ values that we explore~\cite{Hojjati:2012rf}. In the following analysis we consider models with $|f^0_R| \in [10^{-6}, 10^{-4}]$ and $n=1$.

\begin{figure*}[!th]
\begin{center}
\includegraphics[width=0.9\textwidth]{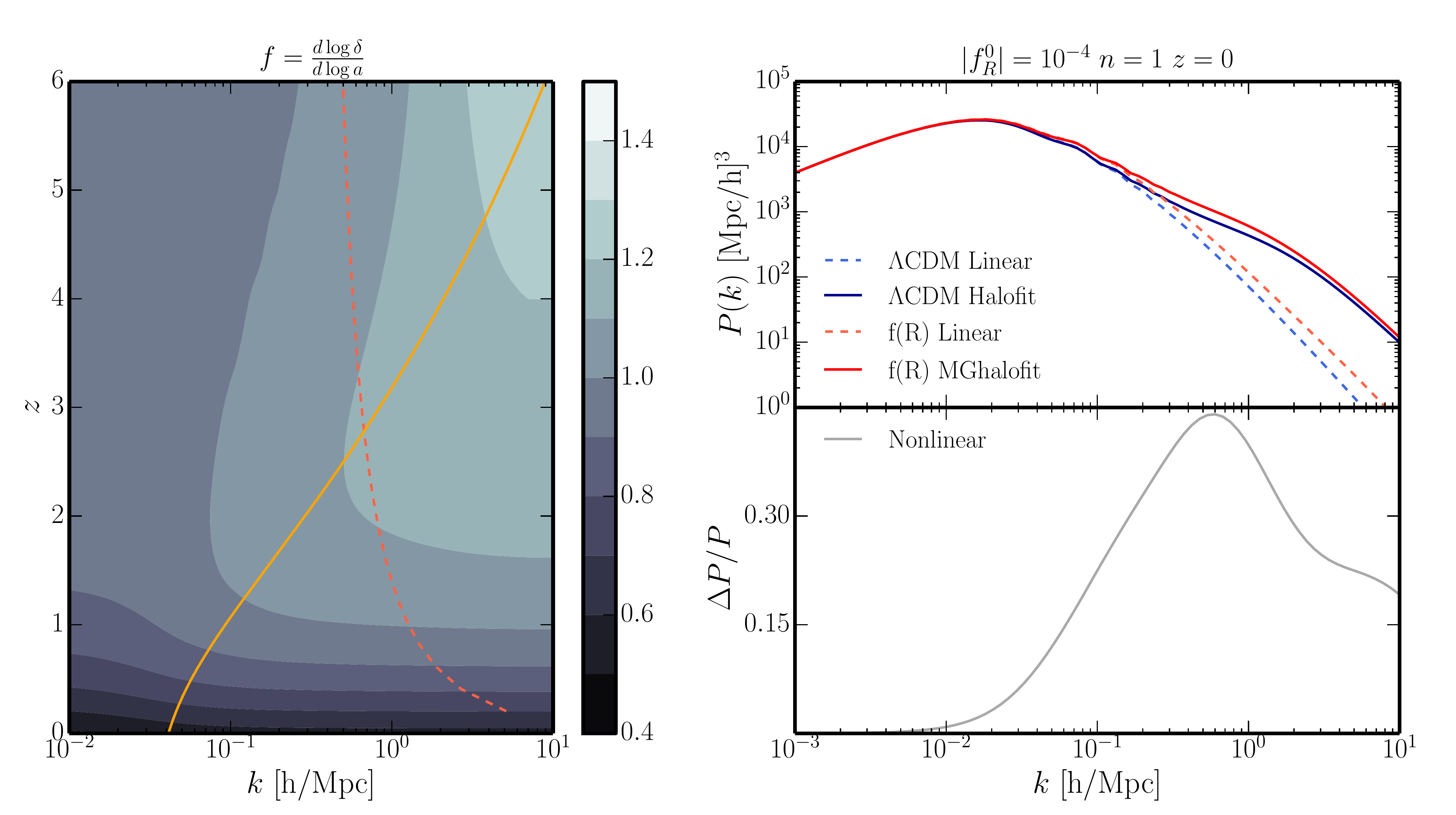}
\caption{Right panel: The linear growth rate $f = d\log{\delta}/d\log{a}$ as a function of scale $k$ and redshift $z$ for the $f(R)$ model with a $\Lambda$CDM background expansion with $|f^0_R|=10^{-4}$ and $n=1$. The characteristic scale dependence of the structure growth is clearly visible: at a given redshift, different scales are characterized by a different growth rate. The red dashed line represents the size-distance relation, $k = \ell/x(z)$, for $\ell = 3000$ while the orange solid line corresponds to the Compton scale $k_C$ associated to the scalaron. Top righ panel: Comparison between matter power spectra in $\Lambda$CDM and $f(R)$ models at redshift $z=0$: linear predictions are shown as dashed lines, while nonlinear ones are shown as solid lines. Nonlinear matter power spectra in $\Lambda$CDM are evaluated using Halofit prescription, while we make use of MGHalofit for the Hu and Sawicki model (with $|f^0_R|=10^{-4}$ and $n=1$). Bottom right panel:  The relative difference between the \emph{nonlinear} matter power spectra in $f(R)$ and $\Lambda$CDM cosmology.}
\label{Fig:growth_pk}
\end{center}
\end{figure*}

\section{Results and Discussion}\label{Sec:Results}
We now proceed to calculate the kSZ  power spectrum for the above models. Before discussing the final results, let us give an overview of the different quantities that contribute to $\Delta^2_B$~(\ref{delta_B_nl}) and how they differ in these models.  In particular, we  discuss how the differences in the dynamics of perturbations for $f(R)$ models reflect in a modified kSZ signal. We also give an overview of other physical phenomena to which the kSZ is sensitive regardless of the theory of gravity under consideration, discussing how they can affect our ability to test gravity with kSZ.

\subsection{Growth history}
Being a weighted integral of $\Delta^2_B$ over the redshift, the kSZ power spectrum probes the cosmic growth history across a wide range of scales and redshifts through the evolution of matter and velocity perturbations~\cite{Shaw:2011sy}. In our analysis of $f(R)$ models, we obtain the linear growth rate $f(z,k)$ using MGCAMB and following the procedure outlined in~\cite{Ma:2013taq}: first we output the density contrast as a function of redshift and scale, and then we evaluate its logarithmic derivative with respect to the scale factor. In the left panel of Fig.~\ref{Fig:growth_pk} we show the growth rate as a function of scale $k$ and redshift $z$ for the Hu and Sawicki model with $|f^0_R|=10^{-4}$: it  can be seen that $f(R)$ models  enhance the growth of matter perturbations on scales smaller than their characteristic length scale, i.e. the scalaron Compton wavelength $\lambda^2_C \sim f_{RR}/(1+f_R)$. On the contrary, the growth in $\Lambda$CDM depends only on time. 

The differences in the growth history show up in the matter power spectrum. In the right panel of Fig.~\ref{Fig:growth_pk}, we plot the present time matter power spectrum, $P_{\delta\delta}(k)$, for $\Lambda$CDM and the Hu-Sawicki $f(R)$ model with $|f^0_R|=10^{-4}$, including both the linear and nonlinear cases. It can be noticed that the scale-dependent growth of $f(R)$ produces an enhancement of the power spectrum on smaller scales, with the effect kicking in at a $k$ proportional to the Compton wave number. The bigger $f^0_R$, the smaller the $k$ at which the enhancement kicks in.

\begin{figure}[!th]
\begin{center}
\includegraphics[width=0.5\textwidth]{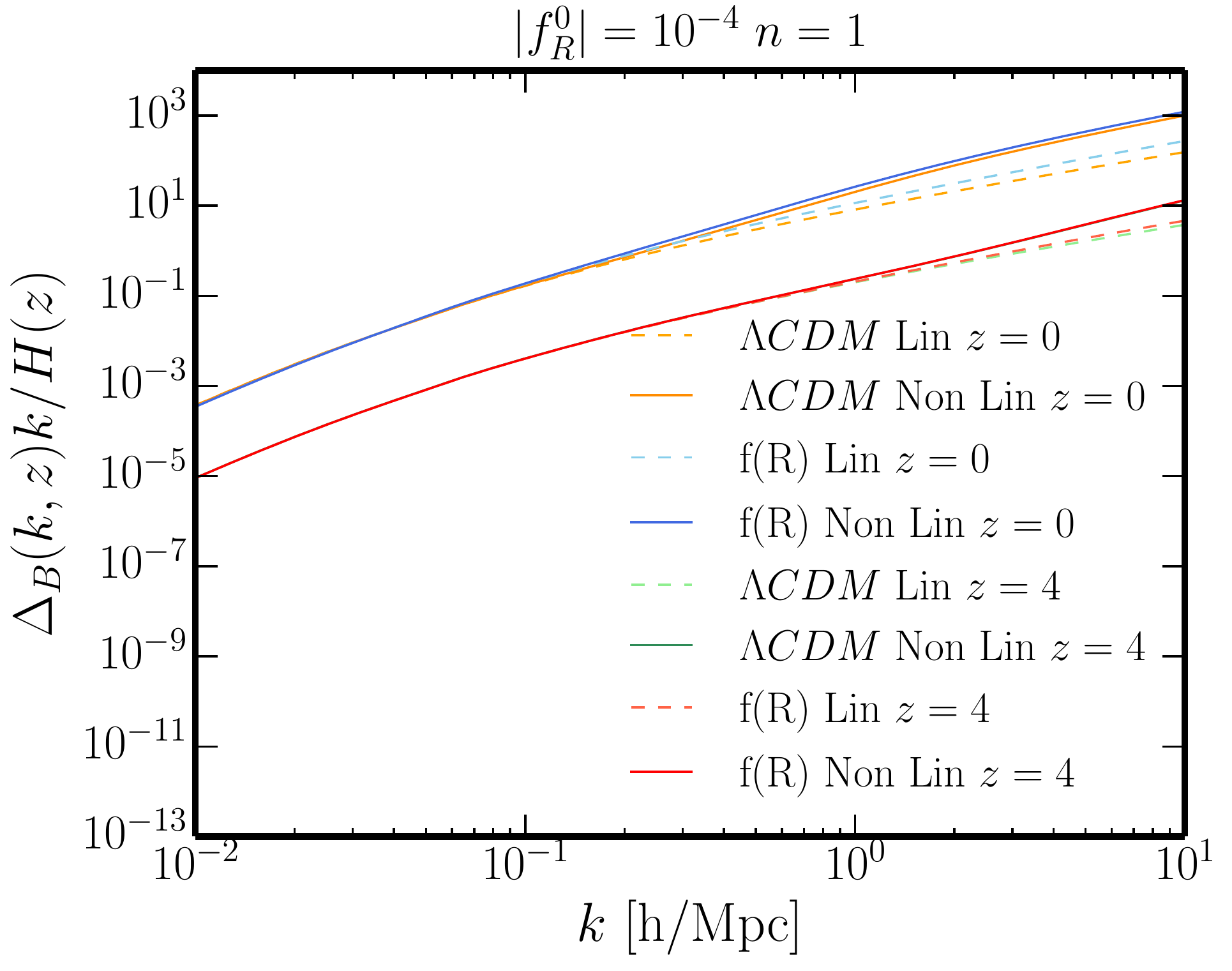}
\caption{Power spectrum of the curl component of the momentum field calculated for the different gravity model, regimes, and redshifts. }
\label{Fig:Delta_b}
\end{center}
\end{figure}

In Fig.~\ref{Fig:Delta_b} we plot the momentum power, $\Delta^2_B$, in terms of the dimensionless quantity $\Delta_Bk/H(z)$ introduced in \cite{Zhang:2003nr,Shaw:2011sy}: dashed lines represent linear regime calculations (OV effect), while solid lines include nonlinear corrections to the matter power spectra (full kSZ effect). 
The plot shows that as cosmic structure evolves over time, the amplitude of the momentum curl component power spectra increases: that is equally true for $f(R)$ and $\Lambda$CDM. However, the modifications of the growth of structure in $f(R)$  imprint a different shape on $\Delta^2_B$ and enhance its power. Nonlinear density fluctuations become relevant when $\Delta_Bk/H(z) \approx 1$: as the Universe evolves, density perturbations exceed unity and they increasingly become important at larger scales \citep{Zhang:2003nr}. 

\subsection{The kSZ signal}
We now move on to understand how the differences at the perturbation level translate into differences in the predicted kSZ angular power spectrum, showing how kSZ measurements can potentially represent a novel test of gravity.

\begin{figure*}[!th]
\begin{center}
\includegraphics[width=0.84\textwidth]{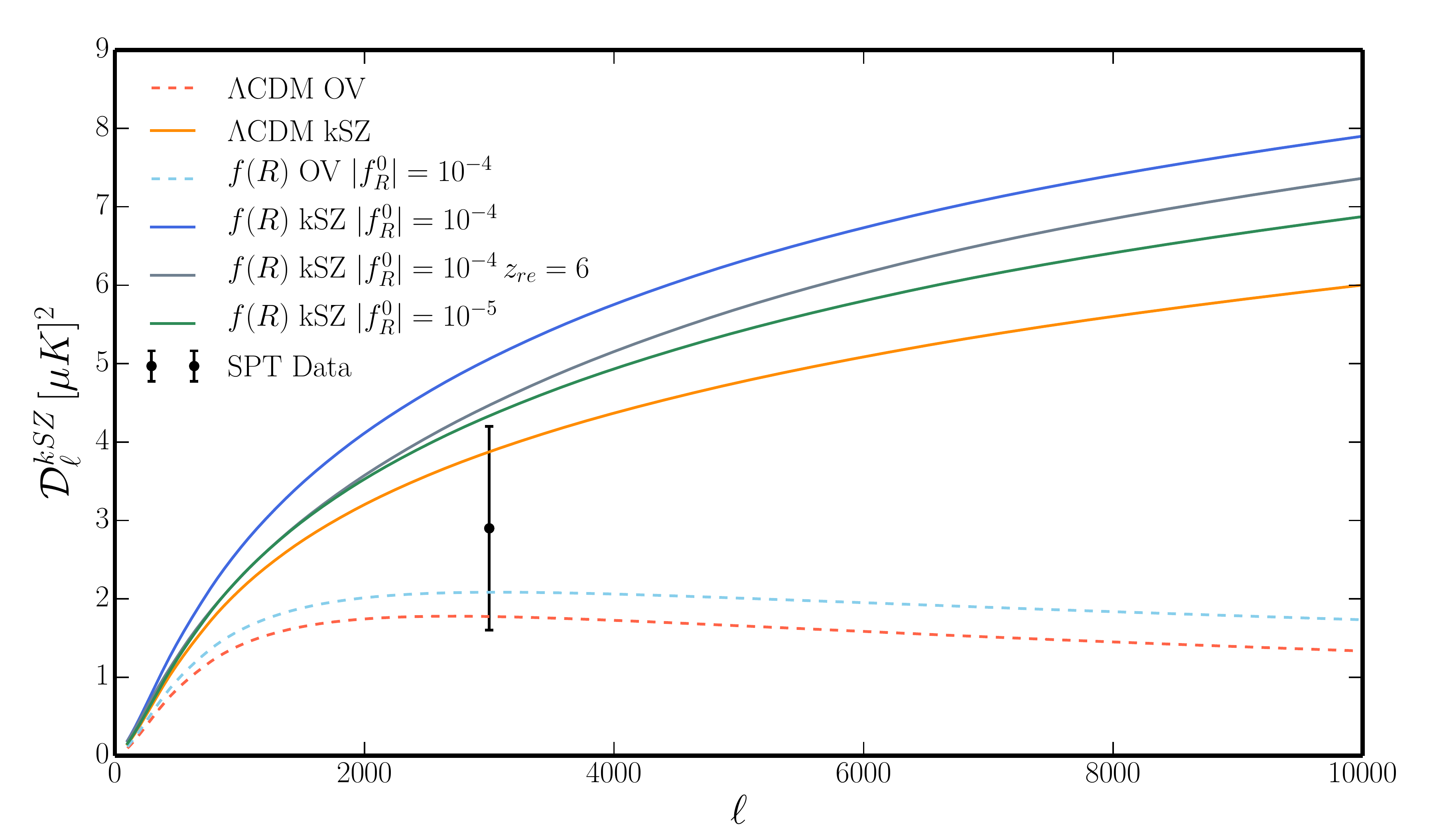}
\caption{The homogeneous kinetic Sunyaev-Zel'dovich power spectrum (solid lines) for standard $\Lambda$CDM and Hu and Sawicki models with $|f^0_R|=\{10^{-5},10^{-4}\}$ and $n=1$ as a function of multipole $\ell$. The dashed lines show the linear predictions, i.e. the OV effect. The black data band power $\mathcal{D}^{\rm kSZ}_{\ell=3000} = 2.9 \pm 1.3 \mu$K$^2$ ($1\sigma$ confidence level) is taken from the South Pole Telescope (SPT) \citep{George:2014oba}). $z_{\rm re} = 9.9$ is assumed except where otherwise stated.} 
\label{Fig:ksz}
\end{center}
\end{figure*}

In Fig.~\ref{Fig:ksz} we show the kSZ angular power spectrum  as a function of multipole $\ell$ in terms of $\mathcal{D}_{\ell} \equiv \ell(\ell+1)C^{\rm kSZ}_{\ell}/2\pi$. Full kSZ theory spectra are plotted as solid lines while OV calculations are shown as dashed lines. We plot results for the $\Lambda$CDM scenario and for two Hu and Sawicki models with $|f^0_R|=\{10^{-5},10^{-4}\}$. The comparison between dashed and solid lines for a fixed model provides an estimate of the impact of nonlinearities in structure formation to the kSZ power spectrum, as also shown in \cite{Shaw:2011sy,Ma:2013taq}. Nonlinear structure boosts the $\Lambda$CDM homogeneous kSZ power by a factor $\approx 2.2$ at $\ell=3000$. On top of nonlinearities, in $f(R)$ cosmology the scale-dependent growth rate enhances the power spectrum by $\approx 2.5$ ($2.4$) for $|f^0_R|=10^{-4}$ ($10^{-5}$). The net effect of modifying gravity is to enlarge the expected kSZ signal: the $C^{\rm kSZ}_{\ell}$ amplitude increases as $|f^0_R|$ becomes larger (i.e. chameleon mechanism is less efficient). Note that the $\ell$-dependence is not dramatically altered. When comparing full kSZ power for different gravity models, we find $C^{\rm kSZ}_{\ell}(|f^0_R|=10^{-4})$ is approximately 30\% larger than the $\Lambda$CDM values, while $C^{\rm kSZ}_{\ell}(|f^0_R|=10^{-5})$ is $\approx11\%$ larger than the $\Lambda$CDM one.

\subsection{\textbf{Fitting formula}} 
\begin{figure}[!th]
\begin{center}
\includegraphics[width=0.5\textwidth]{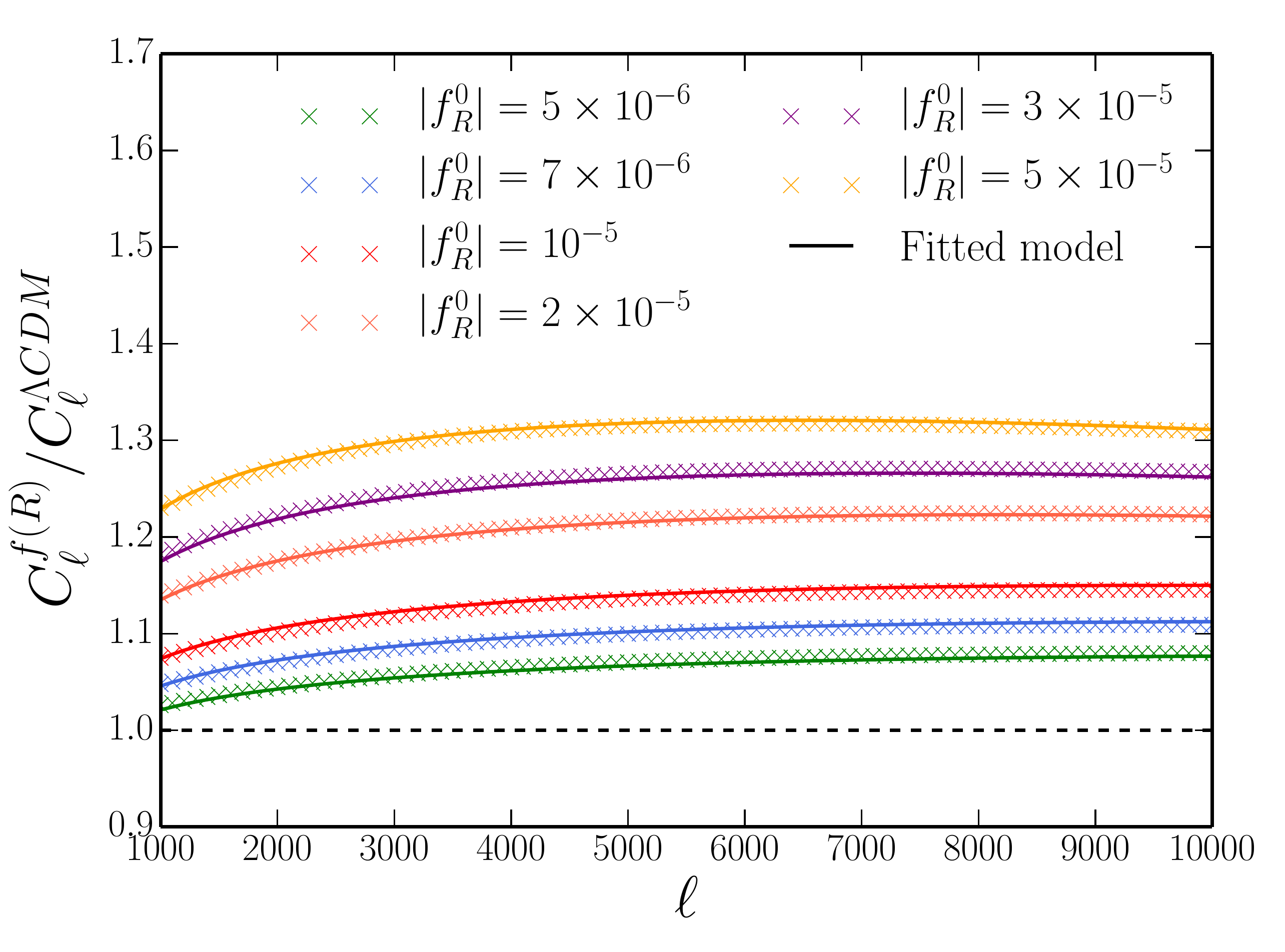}
\caption{ $C_{\ell}^{\rm kSZ}$'s ratio between $\Lambda$CDM and Hu \& Sawicki model for different values of $|f_R^0|$ with the fitting formula introduced in Eq.~\ref{fitting_formula}. Notice that the fit has been done using $C_{\ell}$'s between $\ell \in [10^3,10^4]$ output for six $|f^0_R|$ values in the range $[5 \times 10^{-6},5\times10^{-5}]$.} 
\label{Fig:cl_ratio}
\end{center}
\end{figure}

As noted in~\cite{Ma:2013taq}, the numerical evaluation of $C^{\rm kSZ}_{\ell}$'s is computationally expensive and makes the exploration of the parameter space  through Markov chain Monte Carlo (MCMC)  rather unfeasible. To this end, we develop a fitting formula to predict the kSZ angular power spectrum for the Hu-Sawicki $f(R)$ model with $n=1$. We consider the power spectra ratio $C^{f(R)}_{\ell}/C^{\Lambda CDM}_{\ell}$ for several values of  $|f^0_R|$ in the representative range $[5\times 10^{-6},5\times10^{-5}]$ and perform a fit via the following formula
\be\label{fitting_formula}
\frac{C^{f(R)}_{\ell}(|f^0_R|)}{C^{\Lambda CDM}_{\ell}} = A \Bigl( \frac{\ell}{1000} \Bigr)^{B} \exp{\Bigl(-C\frac{\ell}{1000} \Bigr)},
\ee
where $A$ depends on $|f^0_R|$ through $a_0 + a_1\log_{10}(|f^0_R|) + a_2 \log^2_{10}(|f^0_R|)$ (same dependence for $B$ and $C$). The outcome of this fitting procedure gives: 
\be
\begin{split}\label{fit_Cl}
A(|f^0_R|) &= 3.308 + 0.603\log_{10}(|f^0_R|) + 0.034 \log^2_
{10}(|f^0_R|) \\
B(|f^0_R|) &= -0.261 - 0.169\log_{10}(|f^0_R|) - 0.021 \log^2_{10}(|f^0_R|) \\
C(|f^0_R|) &= 0.182 + 0.044\log_{10}(|f^0_R|) + 0.002 \log^2_{10}(|f^0_R|).
\end{split} 
\ee
As can be noticed in Fig.~\ref{Fig:cl_ratio}, the above curves are a very good fit. Hence  Eq.~(\ref{fitting_formula}) provides an accurate way of modeling the kSZ spectrum in the $f(R)$ models under consideration, allowing one to explore the parameter space in a much faster way. Let us notice that the fitting parameters in~(\ref{fit_Cl}), and possibly the optimal fitting curve, may change if a different range of values for $|f_R^0|$ is considered. 
 
\subsection{Caveats \emph{et al.}}\label{Sec:caveats}
The focus of our study is the impact of modifications of gravity on the kSZ signal. However, since we aim at determining whether the kSZ can be a useful probe for modified gravity, in this section we review astrophysical and cosmological effects that could be degenerated with the main signatures we are after, as well as  limitations of the theoretical modeling.  

Reionization redshift - The amplitude of the kSZ power spectrum depends on the redshift $z_{\rm re}$ at which reionization occurs through Eq.~\ref{cl_ksz}. If reionization ends at higher redshifts, the integral which sources the kSZ power picks up more signal, hence it increases. We find that by fixing $z_{\rm re} = 6$, $C^{\rm kSZ}_{\ell=3000}$ is decreased by approximately 15\% and 11\% with respect to the baseline $\Lambda$CDM and Hu and Sawicki ($|f^0_R|=10^{-4}$) results.

Helium reionization - So far, we have assumed that helium remains neutral throughout cosmic history ($\chi = 0.86$), although in a more realistic model helium would be singly ionized ($\chi = 0.93$) between $3 < z \le z_{\rm re}$ and doubly ionized ($\chi=1$) for $z \le 3$. The magnitude of $C^{\rm kSZ}_{\ell}$ scales as the square of the ionization fraction (plus integrated dependence through the optical depth $\tau$), so we expect that helium reionization would boost the power spectrum.

\begin{figure*}[!th]
\begin{center}
\includegraphics[width=0.9\textwidth]{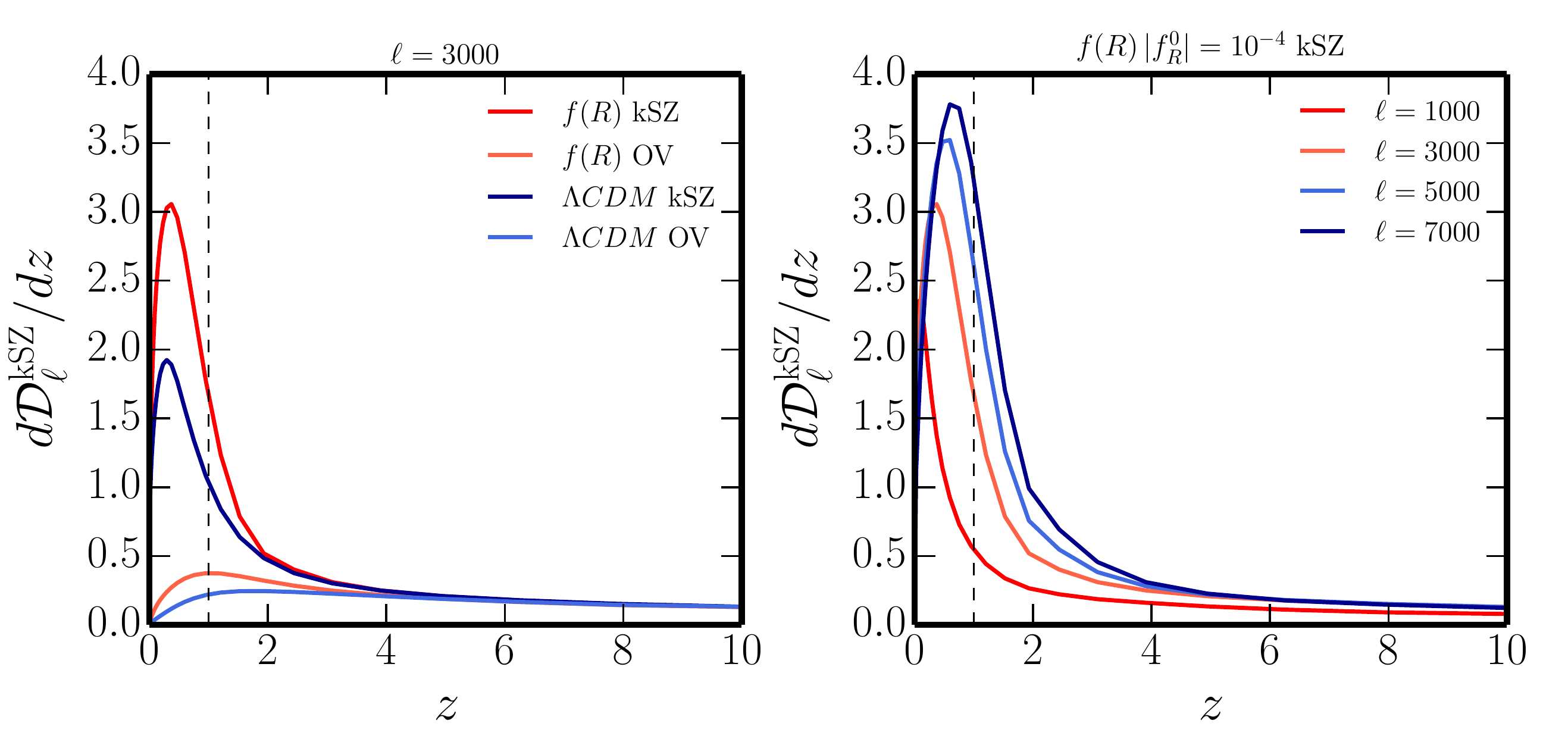}
\caption{\emph{Left panel:} the contribution to the kSZ power as function of redshift at $\ell = 3000$ shown for $\Lambda$CDM and Hu \& Sawicki model with $|f^0_R|=10^{-4}$ and $n=1$. A comparison between the full kSZ and OV effect is also provided. \emph{Right panel:} redshift contribution to kSZ power in the $f(R)$ model for different angular scales. Dashed line marks $z=1$ in both panels.}
\label{Fig:dDl_ksz}
\end{center}
\end{figure*}

Applicability of MGHalofit - A fundamental ingredient needed to predict the kSZ power is the matter power spectrum, especially the nonlinear counterpart: as stated in Sec.~\ref{Sec:Models}, in our analysis we make use of Halofit for the $\Lambda$CDM case (as done in \cite{Shaw:2011sy,Ma:2013taq}) and MGHalofit for the $f(R)$ model. The accuracy level of the fitting formula to calculate the nonlinear matter power spectrum is crucial if one wants to test gravity with kSZ effect. We recall that MGHalofit works for an arbitrary $|f^0_R| \in [10^{-6},10^{-4}]$ below redshift $z = 1$ (reaching an accuracy of 6\% and 12\% at $k \le 1 \,h$/Mpc and $k \in (1, 10]\, h$/Mpc)\footnote{Note that the standard Halofit accuracy is below 5\% ($k \le 1 h$/Mpc) and 10\% ($k \in (1, 10] h$/Mpc) at $z \le 3$ \citep{halofit2}, so that previous analytical $\Lambda$CDM calculations are similarly affected by the precision of the fitting formula for matter nonlinearities.}. While it is nontrivial to address quantitatively and self consistently this issue, we expect nonlinearities to be less important at redshift $z\gtrsim1$. In the left panel of Fig.~\ref{Fig:dDl_ksz} we show the differential contribution to the kSZ power of redshift slices in the range $0 \le z \le z_{\rm re}$ at $\ell = 3000$ for the different cases studied in our analysis. The difference between the OV and kSZ lines (within the same gravity scenario) gives an estimate of the kSZ power enhancement due to nonlinear evolution: moreover, we can see that in the full kSZ case, the bulk of the signal is sourced by structure at $z \lesssim 1$ [especially for the $f(R)$ model]. The right panel of Fig.~\ref{Fig:dDl_ksz} shows the differential contribution $d\mathcal{D}^{\rm kSZ}_{\ell}/dz$ for the $f(R)$ kSZ case with $|f^0_R|=10^{-4}$ at different $\ell$'s: as the multipole becomes larger, the differential redshift distribution peak shifts toward higher $z$ (but always smaller than 1).

Gastrophysics -  Shaw \emph{et al.}~\cite{Shaw:2011sy} have shown that the effect of the baryon physics, i.e. cooling and star formation, is to reduce the gas density in halos, hence hindering the kSZ power boost due to nonlinear density fluctuations. The authors find a reduction of $C_{\ell}^{\rm kSZ}$ at all angular scales; in particular the power in the CSF model is reduced by $\approx 30\%$ at $\ell=3000$ with respect to the model without baryon physics. However, they argue that their radiative simulation suffers from the overcooling problem, so that the measured kSZ power is likely to be underestimated.

Patchy Reionization -
What we have modeled so far is the homogeneous kSZ signal which is sourced in the postreionization epoch: however, the patchy kSZ contribution should be added on top of that. The precise patchy kSZ power spectrum amplitude and $\ell$-shape depend at first order on the details of reionization, i.e. its time and duration \citep{2005ApJ...630..643M,2005ApJ...630..657Z,Iliev:2006zz}. In particular, the patchy kSZ signal has a different spectral shape with respect to the homogeneous one and peaks on multipoles between $\ell \approx 2000 - 8000$, roughly corresponding to the typical angular size of ionized regions at the reionization redshift \citep{Iliev:2006zz}. Patchy kSZ can allow us to place constraints on the reionization physics, details of which are not yet well understood, assuming a good knowledge and modeling of the homogeneous signal. To this end, 21-cm fluctuation power spectra and its cross-correlation with the CMB anisotropies can shed light on the details of reionization and help disentangle between different kSZ contributions~\cite{Jelic:2015cia}.

\section{Observational Outlook}\label{Sec:kSZ_obs}
The kSZ effect was detected observationally for the first time by~\cite{hand_ksz}, applying the pairwise momentum estimator to data from the Atacama Cosmology Telescope (ACT) and the Baryon Oscillation Spectroscopic Survey (BOSS). Recently some authors have performed  an analysis of  constraints on dark energy and modifications to gravity achievable with  the mean pairwise velocity of clusters estimator~\cite{Mueller:2014nsa}. Here, we focus on measurements of the kSZ angular power spectrum obtained from CMB surveys. At high multipoles ($\ell \gtrsim 3000$), there are several sources contributing to the secondary anisotropies of the observed CMB temperature: a number of galactic and extragalactic astrophysical foregrounds, such as the cosmic infrared background (CIB), thermal Sunyeav-Zel'dovich (tSZ) and kSZ, radio galaxies, synchrotron and dust emission \footnote{The effect of weak gravitational lensing is present (but subdominant) at these angular scales and it gently smoothes the CMB peaks.} . Multiband observations are fundamental: the different frequency scalings of these foregrounds can help in separating out the primordial CMB contributions. However, the only contribution that is spectrally degenerate with the primary CMB is the kSZ emission which has a blackbody spectrum. 

To the best of our knowledge, the latest observational constraints on the kSZ power spectrum were reported in~\cite{George:2014oba}  from the  combination of  the 95, 150, and 220 GHz channel data from the SPT. By jointly fitting the data for the tSZ and kSZ templates (assuming equal power at $\ell=3000$ from the homogenous (CSF) and patchy contributions), it is found that $\mathcal{D}^{\rm kSZ}_{3000} < 5.4 \mu$K$^2$ at 95\% C.L.; when tSZ bispectrum information is added, the derived constraint on the kSZ amplitude is  $\mathcal{D}^{\rm kSZ}_{3000} = 2.9 \pm 1.3\mu$K$^2$ (this data point is shown in Fig.~\ref{Fig:ksz}). Comparing with our findings, Fig.~\ref{Fig:ksz}, we can see that current kSZ measurements would already have  some constraining power. More importantly, we expect kSZ measurements   from the ongoing and upcoming CMB surveys to provide a novel complementary probe of gravity on cosmological scales. In particular, it will be essential to rely on arcminute-scale resolution and high sensitivity CMB data since the instrument capability to detect kSZ signal degrades substantially when enlarging the beam size, as noted in~\cite{Calabrese:2014gwa}. In the latter paper, the authors investigate the possibility of using the small-scale polarization information to constrain primordial cosmology in order to remove the primary CMB from temperature measurements, hence isolating the kSZ contribution (assuming an efficient foreground cleaning from multifrequency channels).
Another approach proposed to detect the kSZ signal is to infer the peculiar velocity field $\mathbf{\hat{v}}$ from the observed galaxy number overdensity\footnote{By solving the linearized continuity equation in redshift space.} $\delta_g$ and to stack the CMB temperature maps at the location of each halo, weighted by the reconstructed  $\bf{\hat{v}}$ field. Recent analysis exploiting this technique has been reported in~\cite{Ade:2015lza,Schaan:2015uaa}.

\section{Conclusions}\label{Sec:concl}
We have revisited the kSZ effect in the context of modified theories of gravity that approach the phenomenon of cosmic acceleration. In particular, we have focused on the class of $f(R)$ models introduced by Hu and Sawicki in~\cite{Hu:2007nk}. We have found that, as expected, the kSZ effect is particularly sensitive to modifications of the growth rate of structure offering,  in principle, an interesting complementary probe  for modified gravity. Interestingly, we find the kSZ signal to be more sensitive to modifications of the dynamics of cosmological perturbations than to those of the expansion history. 

As we have discussed in Sec.~\ref{Sec:caveats}, there are several assumptions and caveats in the modeling of the kSZ effect that could hinder its power in constraining modified theories of gravity. We have given a detailed overview of these, elaborating on possible ways of overcoming them, also in light of future experiments. Finally, as this is always the case when the growth rate of structure plays an important role, we expect a degeneracy between the modifications of the kSZ induced by $f(R)$ gravity, that we have discussed, and those that would be induced by massive neutrinos. As part of future work, it would certainly be interesting to explore this latter aspect, as well as to use N-body simulations to get  the full nonlinear velocity and density power spectra.  

Upcoming high resolution CMB surveys will carry out multifrequency observations in future years \citep{Calabrese:2014gwa}, allowing for a reconstruction of the small-scale temperature power spectrum and providing a unique window on the cosmic growth history as well as  processes of the epoch of reionization. Given the wide range of modified gravity observational probes and their associated systematics, cross-correlation methods will prove to be robust and complementary tools to determine gravity properties and constrain its behavior over cosmic time. 

\begin{acknowledgments}
We are grateful to Bin Hu, Yin-Zhe Ma and Gong-Bo Zhao for useful discussions. F.B. acknowledges partial support from the INFN-INDARK initiative.  A.S. acknowledges support from The Netherlands Organization for Scientific Research (NWO), and also from the D-ITP consortium, a program of the NWO that is funded by the Dutch Ministry of Education, Culture and Science (OCW). 
\end{acknowledgments}

\end{document}